\documentclass[12pt,twocolumn,english]{iopart}
\pdfoutput=1
\usepackage{mathptmx}
\usepackage[T1]{fontenc}
\usepackage[latin9]{inputenc}
\usepackage{color}
\usepackage{babel}
\usepackage{amstext}
\usepackage{graphicx}
\usepackage[unicode=true,pdfusetitle,
 bookmarks=true,bookmarksnumbered=false,bookmarksopen=false,
 breaklinks=true,pdfborder={0 0 0},pdfborderstyle={},backref=false,colorlinks=true]
 {hyperref}
\hypersetup{
 linkcolor=blue,citecolor=blue,urlcolor=blue}

\makeatletter

\providecommand{\tabularnewline}{\\}

\usepackage{iopams}

\usepackage{cite}

\@ifundefined{showcaptionsetup}{}{%
 \PassOptionsToPackage{caption=false}{subfig}}
\usepackage{subfig}
\makeatother

\begin{document}
\title{Dynamics of extended Schelling models}
\author{André P. Vieira}
\address{Instituto de Física, Universidade de São Paulo, 05508-090, São Paulo,
SP, Brazil}
\author{Eric Goles}
\address{Facultad de Ingeniería y Ciencias, Universidad Adolfo Ibáñez, Avenida
Diagonal las Torres 2640, Peñalolén, Santiago, Chile}
\author{Hans J. Herrmann}
\address{Departamento de Física, Universidade Federal do Ceará, 60451-970 Fortaleza,
CE, Brazil}
\address{ESPCI, CNRS UMR 7636 - Laboratoire PMMH, 75005 Paris, France}
\begin{abstract}
We explore extensions of Schelling's model of social dynamics, in
which two types of agents live on a checkerboard lattice and move
in order to optimize their own satisfaction, which depends on how
many agents among their neighbors are of their same type. For each
number $n$ of same-type nearest neighbors we independently assign
a binary satisfaction variable $s_{k}$ which is equal to one only
if the agent is satisfied with that condition, and is equal to zero
otherwise. This defines 32 different satisfaction rules, which we
investigate in detail, focusing on pattern formation and measuring
segregation with the help of an ``energy'' function which is related
to the number of neighboring agents of different types and plays no
role in the dynamics. We consider the checkerboard lattice to be fully
occupied and the dynamics consists of switching the locations of randomly
selected unsatisfied agents of opposite types. We show that, starting
from a random distribution of agents, only a small number of rules
lead to (nearly) fully segregated patterns in the long run, with many
rules leading to chaotic steady-state behavior. Nevertheless, other
interesting patterns may also be dynamically generated, such as ``anti-segregated''
patterns as well as patterns resembling sponges.
\end{abstract}
\maketitle

\section{Introduction}

Aiming at studying racial segregation in American cities, Schelling
formulated one the first mathematical models of social agents around
50 years ago \cite{Schelling1969,Schelling1971}. The spirit of Schelling's
model can be summarized by the presence of two types of agents occupying
the sites of a network and able to move seeking to optimize their
satisfaction, which is determined by how many agents of their same
type are located in their neighborhood. As shown by Schelling, the
model predicts that only a minor preference of the agents for neighbors
of their same type leads to segregation. 

Apart from many applications to study segregation in other fields
such as biology and economy (see e.g. Ref. \cite{Rogers2011} and
references therein), as well as to determine the robustness of Schelling's
model predictions under a change of network topology \cite{Henry2011,Banos2012},
different versions of the model have attracted the attention of physicists
for the last 15 years (see e.g. Refs. \cite{Vinkovic2006,Stauffer2007,Dallasta2008,Gauvin2009,Grauwin2009,Gauvin2010,Lemoy2011,Goles2011,Rogers2011,Rogers2012,Albano2012,Jensen2018}),
following a similar interest in investigating social dynamics using
a variety of statistical-physics models \cite{Ben-Naim1996,Sood2005,Holme2006,Castellano2009,Teza2019}.
Due to the fact that the dynamics of such models is determined by
the optimization of individual variables rather than collective variables
such as the energy, the methods of equilibrium statistical mechanics
are hardly useful, and mostly computer simulations are employed. In
order to make analytical progress in investigating Schelling-like
models, one has to look at coarse-grained versions \cite{Grauwin2009,Rogers2012,Jensen2018,Durrett2014}
involving extended neighborhoods or to perform more radical approximations
such as making assumptions about the long-time behavior \cite{Dallasta2008}
or imposing that each agent has a single neighbor \cite{Rogers2011}. 

Most approaches based on Schelling's model focus on appearance of
segregation, varying the fraction of vacant sites or the minimum tolerated
fraction of neighboring agents of the same type. In contrast, here
we do not restrain ourselves to cases directly inspired by the desire
to model the appearance of segregation in social systems, but rather,
inspired by nonequilibrium statistical-physics models \cite{Marro2005,Henkel2008},
we explore the dynamic variability of extensions of Schelling's model
in which for each number $n$ of same-type neighbors we independently
assign a binary satisfaction variable $s_{k}$ which is equal to one
only if the agent is satisfied with that condition, and is equal to
zero otherwise. This defines 32 distinct satisfaction rules, which
we investigate in detail, focusing on pattern formation and measuring
segregation with the help of an ``energy'' function which is related
to the number of neighboring agents of different types and plays no
role in the dynamics. The model is defined on a square lattice with
no vacancies and a given agent interacts with the 4 agents in its
von-Neumann neighborhood, rather than with the 8 agents in its Moore
neighborhood, as in Schelling's original model.  The dynamics consists
of switching the locations of randomly selected unsatisfied agents
of opposite types. Although most of our results are obtained through
simulations, we also provide a few analytical estimates. 

We show that, starting from a random distribution of agents, only
a small number of rules lead to (nearly) fully segregated patterns
in the long run, with many rules leading to chaotic steady-state behavior.
Nevertheless, other interesting patterns may also be dynamically generated,
such as ``anti-segregated'' patterns as well as patterns resembling
sponges. A crucial role in the dynamics is played by the existence
of fixed points, which are equivalent to the absorbing states familiar
in the nonequilibrium statistical-physics literature \cite{Marro2005,Henkel2008}.
This is detailed in the next section.

\section{The models}

\label{sec:Extended-Schelling-models}We assume that two types of
agents --- which we call ``blue'' (or type 0) and ``red'' (or
type 1) agents --- fully occupy the sites of a $L\times L$ square
lattice (subject to periodic boundary conditions), with half of the
sites randomly occupied by each type of agent. Each agent interacts
with the 4 agents in its von-Neumann neighborhood. The satisfaction
of each agent depends on the number of agents of its same type in
its neighborhood, so that the model would correspond to a (asynchronous)
totalistic automaton \cite{Wolfram1983}. A parameter of the model
is the five-digit binary number $s=s_{0}s_{1}s_{2}s_{3}s_{4}$, with
$s_{n}=1$ indicating that an agent is satisfied having $n$ neighbors
of its same type, and $s_{n}=0$ otherwise. We take the satisfaction
parameter to be the same for both types of agents, and we refer to
a given satisfaction parameter as defining the satisfaction \emph{rule}
of the dynamics. Schelling's original model of segregation \cite{Schelling1969,Schelling1971}
would be closer to rule $00111$, while rules 00000, 00001, 00011,
00111 and 01111 were also investigated in Ref. \cite{Goles2011}. 

In many formulations of Schelling's model, including his own, the
dynamics prescribes that an unsatisfied agent moves to a vacant site.
As our model has no vacancies, we assume that the basic step of the
dynamics is implemented by randomly selecting two unsatisfied agents,
one of each type, and switching their locations, irrespective of whether
the agents become satisfied in their new locations, and irrespective
of the original distance between the agents. We implement this dynamics
for a maximum of $10^{5}$ Monte Carlo (MC) sweeps, with a single
MC sweep corresponding to $L^{2}$ basic steps. (Keep in mind that
two agents always switch location at each basic step.) In this Section
we investigate the effect of all satisfaction rules on the long-time
behavior of the system. The rule 11111 is trivial, as it does not
allow for the existence of unsatisfied agents, thus having no dynamics,
and will not be further discussed. We are then left with 31 distinct
rules.

We measure time in units of the inverse number of unsatisfied agents,
which means that the time increments between consecutive simulation
steps are nonuniform, being given by
\begin{equation}
\Delta t=\frac{1}{N_{u,1}+N_{u,0}},
\end{equation}
in which $N_{u,c}$ represents the number of unsatisfied agents of
type $c\in\{0,1\}$. We follow the time evolution of the fraction
of unsatisfied agents, 
\begin{equation}
\rho_{u}=\frac{N_{u,1}+N_{u,0}}{L^{2}},
\end{equation}
 and of an ``energy'' function defined as 
\begin{equation}
E=-\sum_{\left\langle i,j\right\rangle }\left(2c_{i}-1\right)\left(2c_{j}-1\right),
\end{equation}
in which the sum is over all neighboring pairs of agents and $c_{i}=0$
($c_{i}=1$) if a blue (red) agent occupies site $i$. The energy
function, which is related to the interface density of Refs. \cite{Dallasta2008,Gauvin2010,Rogers2011,Rogers2012},
has its maximum value ($E_{\text{max}}=2L^{2}$) when the agents arrange
themselves in a checkerboard 1x1 pattern, for which all agents only
have neighbors of the opposite type, while the minimum value ($E_{\text{min}}=-2L^{2}+4L)$
corresponds to complete segregation, in which there are two uniform
domains, each containing all the agents of a given type. Values of
$E$ close to zero indicate the existence of various static or dynamic
local patterns in different regions of the system. For some rules,
along the lines of Ref. \cite{Goles2011}, the energy function can
be shown to be either monotonically nonincreasing or nondecreasing
in time, which is useful in the discussion of the possible long-time
behaviors. We also calculate the survival probability $P_{s}\left(t\right)$,
defined as the average fraction of simulations which do not freeze
(as defined below) before time $t$.

There are three possible cases for the qualitative long-time dynamics,
with finer details to be discussed later: (i) the dynamics freezes
after a finite number of steps, with all agents of at least one type
satisfied and zero survival probability at long times; (ii) a large
fraction of the agents become satisfied after a finite number of steps,
with a small fraction of unsatisfied agents (necessarily of both types)
making the dynamics persist indefinitely, reaching a nonzero survival
probability as $t\rightarrow\infty$; (iii) no agent becomes permanently
satisfied, and the dynamics persists indefinitely in an essentially
chaotic manner, with temporary pockets of satisfied agents, again
with a nonzero long-time survival probability.

For cases (i) and (ii) the behavior is associated with the existence
and stability of ``equilibrium'' states --- fixed points of the
dynamics, in which all agents are satisfied --- corresponding to
a given satisfaction rule. Depending on the rule, there may be a huge
number of fixed points (called ``absorbing states'' in the nonequilibrium
phase transitions literature \cite{Marro2005}), including several
which correspond to irregular arrangements. We list below the simplest
\emph{regular} fixed points (FPs).

\begin{figure}
\begin{centering}
\subfloat[]{\begin{centering}
\includegraphics[width=0.16\textwidth]{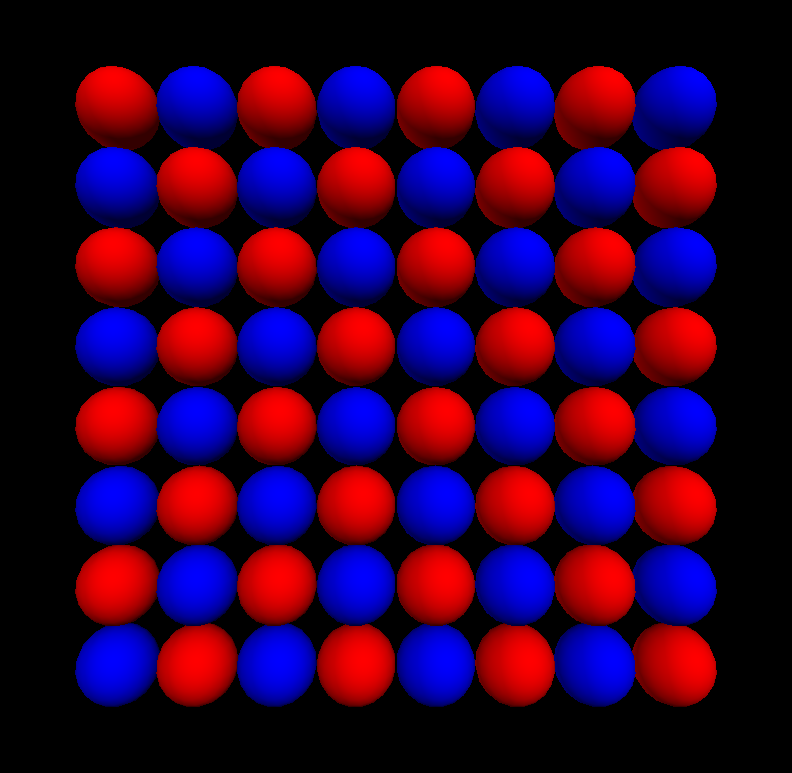}
\par\end{centering}

}\hfill{}\subfloat[]{\begin{centering}
\includegraphics[width=0.16\textwidth]{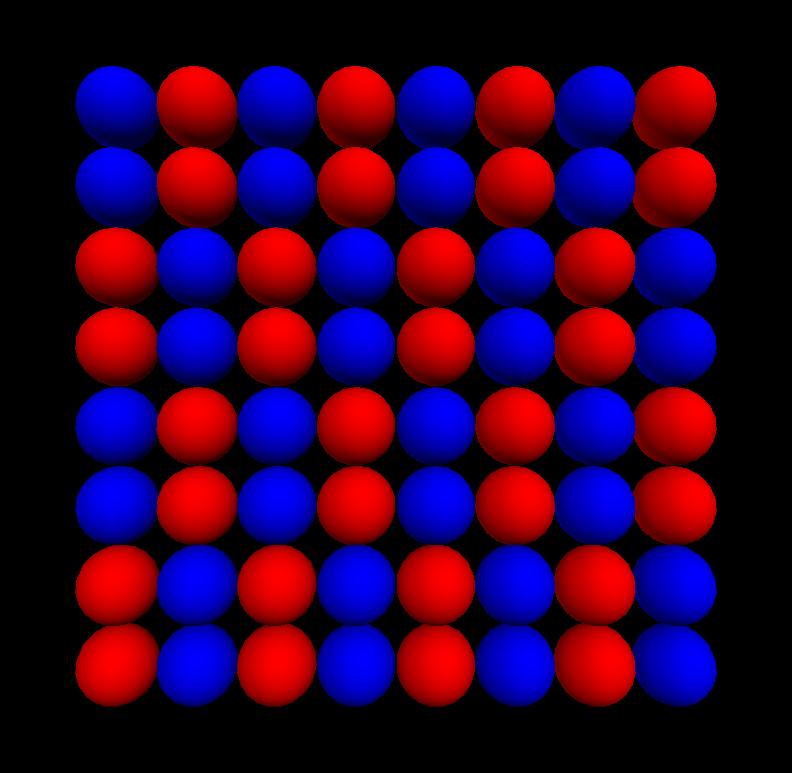}
\par\end{centering}

}\hfill{}\subfloat[]{\begin{centering}
\includegraphics[width=0.16\textwidth]{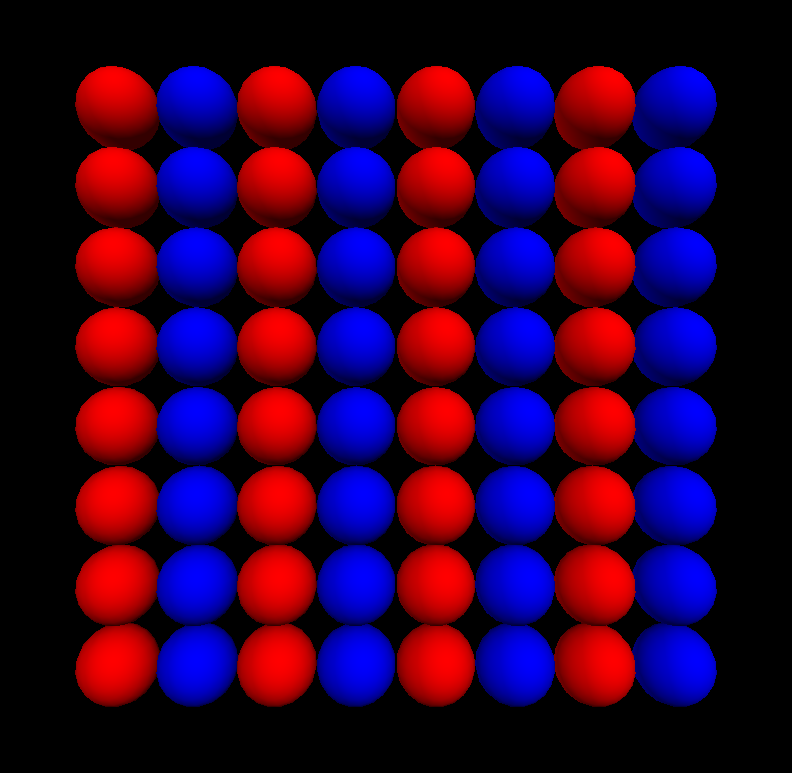}
\par\end{centering}

}\hfill{}\subfloat[]{\begin{centering}
\includegraphics[width=0.16\textwidth]{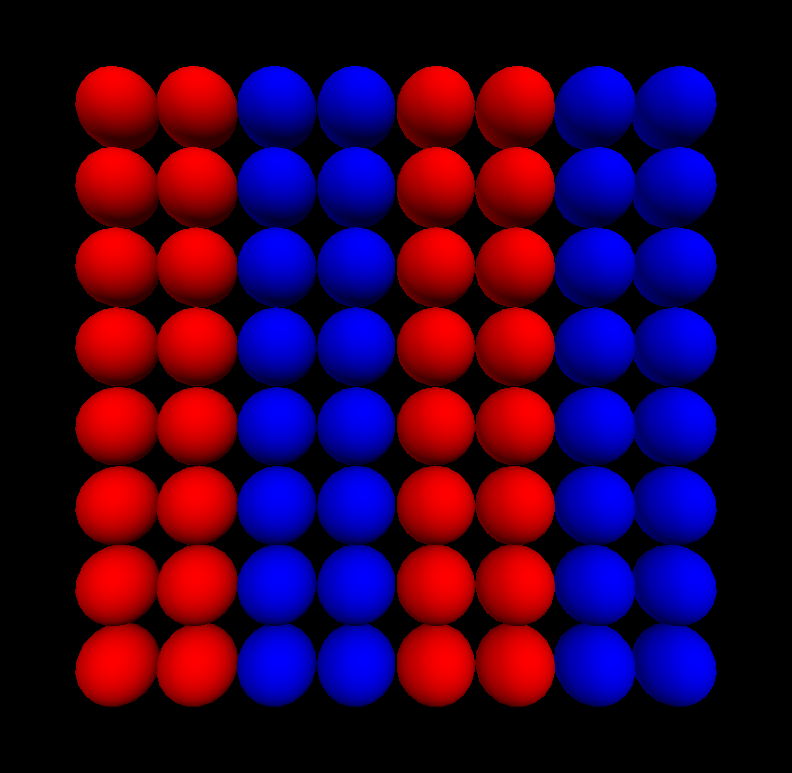}
\par\end{centering}

}\hfill{}\subfloat[]{\begin{centering}
\includegraphics[width=0.16\textwidth]{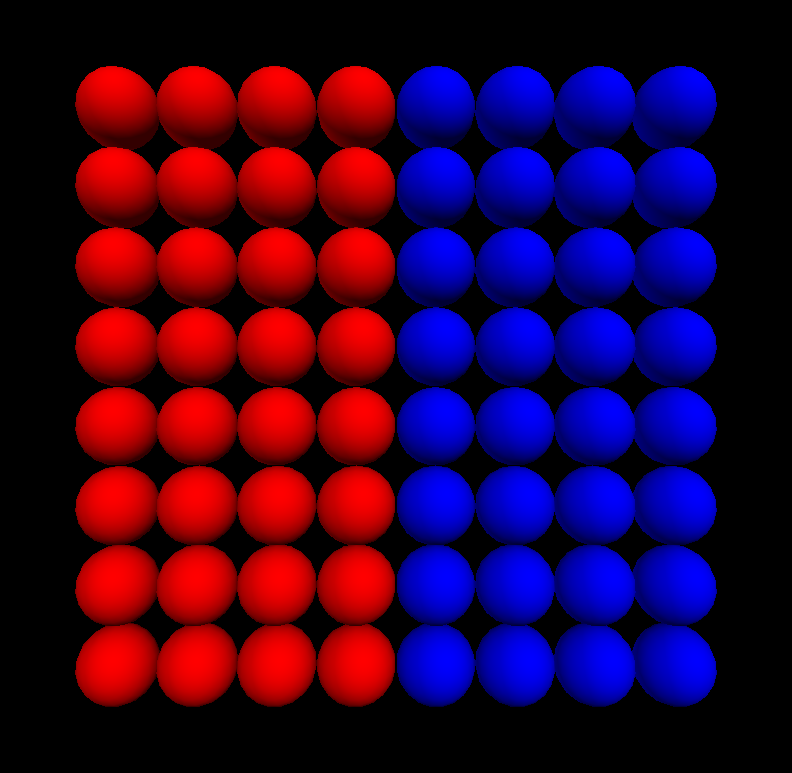}
\par\end{centering}

}\caption{\label{fig:fixedpoints}Some regular fixed points of the rules defined
in Sec. \ref{sec:Extended-Schelling-models}.}
\par\end{centering}
\end{figure}

(1) A checkerboard 1x1 pattern, illustrated in Fig. \ref{fig:fixedpoints}(a),
in which a blue agent is surrounded by four red agents, and vice-versa.
This pattern is a FP for rules of the form $1s_{1}s_{2}s_{3}s_{4}$,
i.e. all rules under which an agent is satisfied having no neighbors
of its same type.

(2) A checkerboard 1x2 pattern, illustrated in Fig. \ref{fig:fixedpoints}(b),
in which a pair of neighboring blue agents is surrounded by six red
agents, and vice-versa. This pattern is a FP for rules of the form
$s_{0}1s_{2}s_{3}s_{4}$, under which an agent is satisfied having
only one neighbor of its same type.

(3) A striped pattern, illustrated in Fig \ref{fig:fixedpoints}(c),
in which a single line (or column) of blue agents is surrounded by
two lines (or columns) of red agents, and vice-versa. This pattern
is a FP for rules of the form $s_{0}s_{1}1s_{3}s_{4}$, under which
an agent is satisfied having two neighbors of its same type.

(4) A double striped pattern, illustrated in Fig \ref{fig:fixedpoints}(d),
in which a double line (or column) of blue agents is surrounded by
two double lines (or columns) of red agents, and vice-versa. This
pattern is a FP for rules of the form $s_{0}s_{1}s_{2}1s_{4}$, under
which an agent is satisfied having three neighbors of its same type.

(5) A fully segregated pattern, illustrated in Fig. \ref{fig:fixedpoints}(e),
in which there are only two uniform domains of agents of each type,
separated by two linear boundaries (due to our choice of periodic
boundary conditions). This pattern is a FP for rules of the form $s_{0}s_{1}s_{2}11$,
under which an agent is satisfied having either three or four neighbors
of its same type.

We investigated the stability of the above FPs with respect to small
perturbations, introducing typically 1 to 8 ``defect'' agents of
each type by switching the positions of randomly chosen agents of
opposite types in order to disturb the FP arrangement. It turns out
that in most cases the dynamics does not take the system arbitrarily
away from the FPs, with the steady-state configurations resembling
the arrangement of agents in the FP, but with a relatively small fraction
of defects, especially when some of these are satisfied at their new
positions. This fraction of defects turns out to be zero, so that
the FP is fully stable, for rules $00011$ {[}around FP (5){]}, $00110$
{[}around FP (4){]}, $00111$ {[}around FP (4){]}, $01100$ {[}around
FP (2){]}, $01110$ {[}around FP (4){]}, $10111$ {[}around FP (4){]},
$11000$ {[}around FP (1){]}, $11010$ {[}around FP (1){]}, $11100$
{[}around FPs (1) and (2){]}, $11101$ {[}around FP (2){]}, and $11110$
{[}around FP (1){]}. In other cases the FPs are fully unstable, and
the steady-state configuration bears no resemblance to the FP arrangement.
This full instability happens for rules $00010$ {[}around FP (4){]},
$00100$ {[}around FP (3){]}, $00101$ {[}around FP (3){]}, $01000$
{[}around FP (2){]}, $01010$ {[}around both FPs (2) and (4){]}, $01011$
{[}only around FP (2){]}, $10000$ {[}around FP (1){]}, $10001$ {[}around
FP (1){]}, $10010$ {[}around both FPs (1) and (4){]}, $10100$ {[}around
both FPs (1) and (3){]}, $10101$ {[}around both FPs (1) and (3){]},
$10110$ {[}only around FP (1){]}, and $11010$ {[}only around FP
(4){]}.

However, we are more interested in whether the above regular FPs can
be reached starting from \emph{random} initial conditions. This happens
only for the checkerboard 1x1 and the fully segregated patterns, but
under most rules for which these FPs are stable the dynamics leads
a finite system to the neighborhood of the FP without ever precisely
reaching it. This is a consequence of the fact that, besides the regular
fixed points listed above, there are also many other competing \emph{irregular}
fixed points, as well as configurations in which defects become trapped,
giving rise to dynamic patterns (blinkers) in which the positions
of unsatisfied agents alternate between a few positions.

\begin{figure}
\begin{centering}
\subfloat[]{\begin{centering}
\includegraphics[width=0.7\columnwidth]{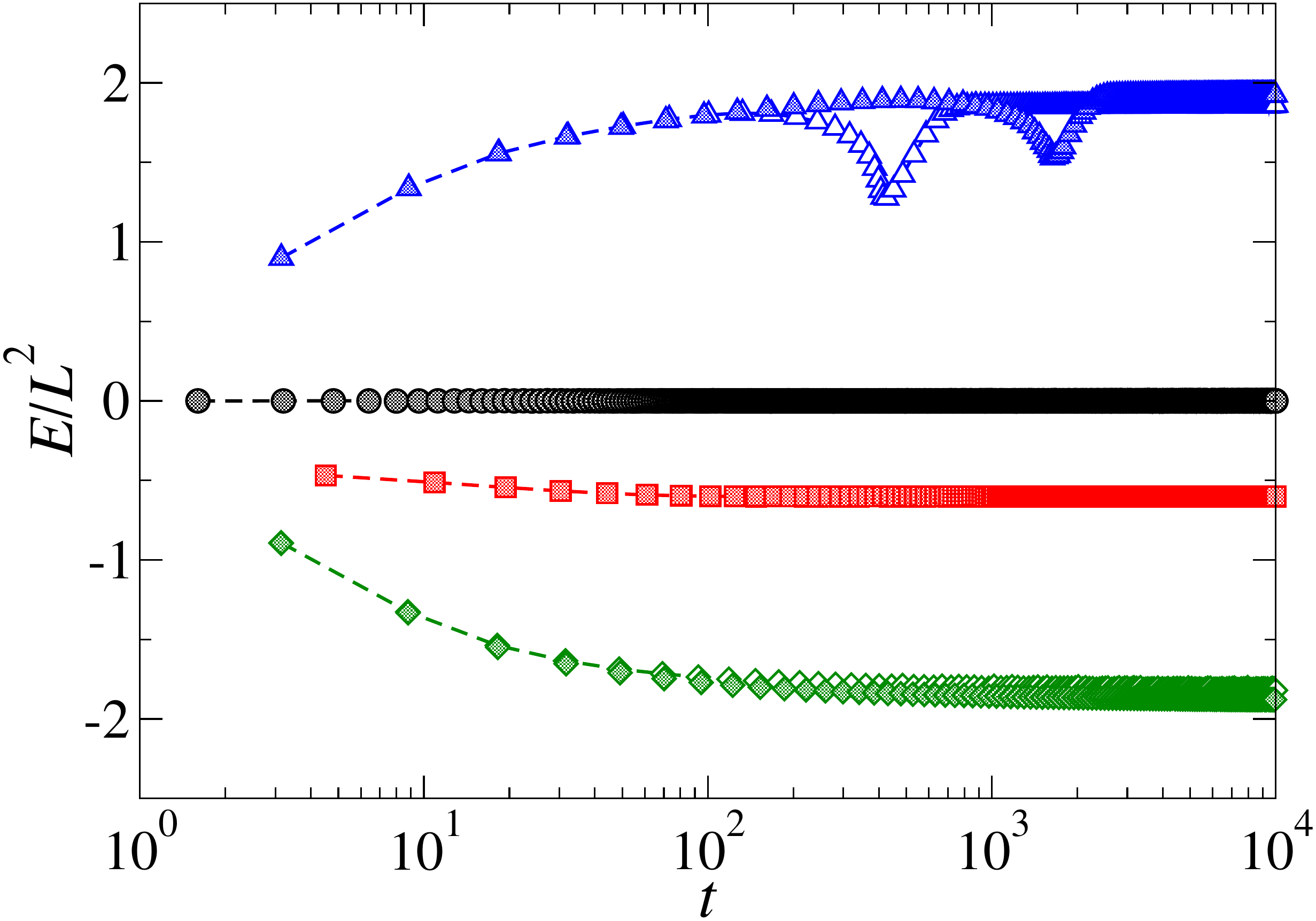}
\par\end{centering}
}
\par\end{centering}
\centering{}\subfloat[]{\begin{centering}
\includegraphics[width=0.7\columnwidth]{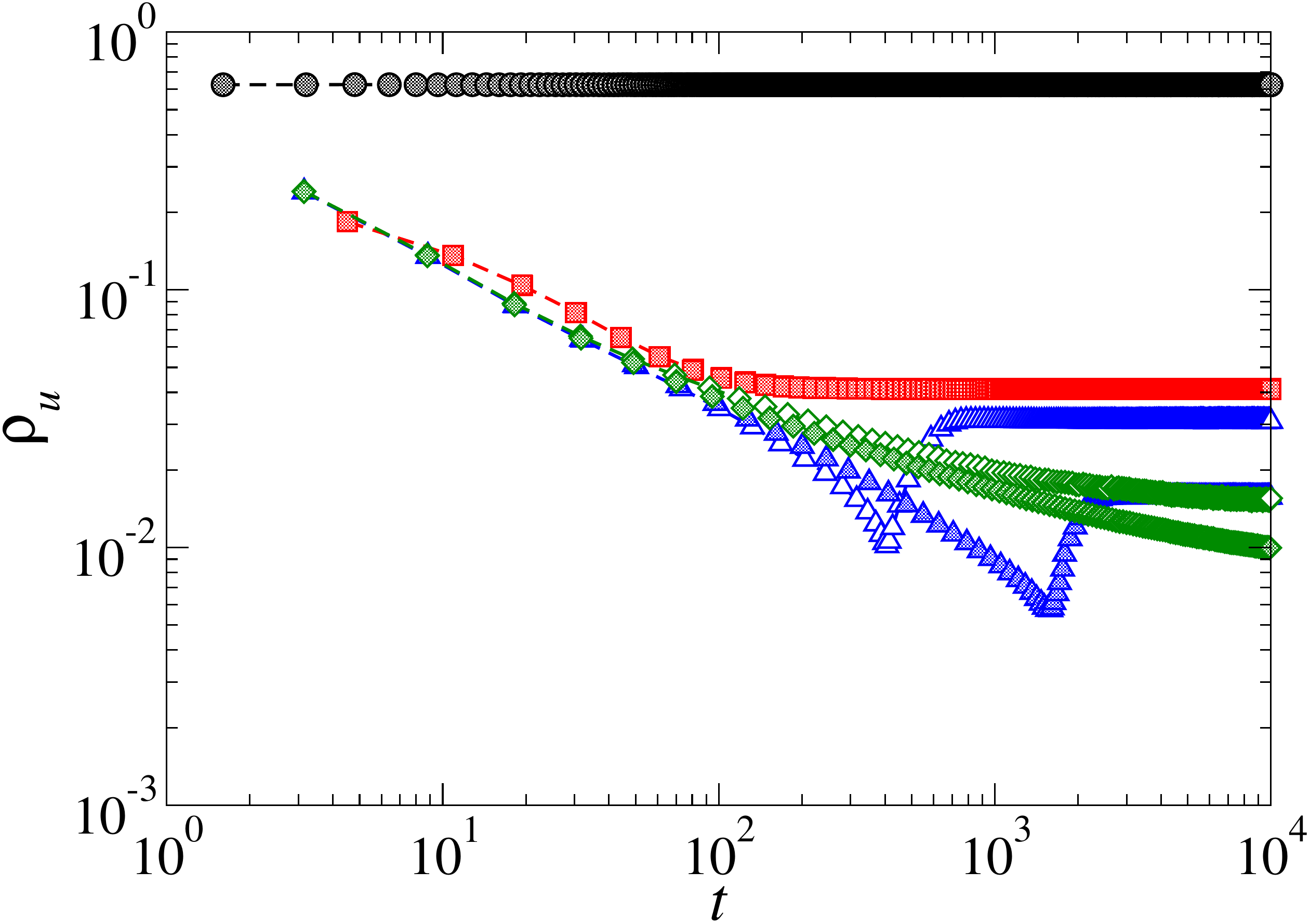}
\par\end{centering}
}\caption{\label{fig:ex_energ_dens}Time dependence of (a) the average energy
function per agent and (b) the average fraction of unsatisfied agents
for rules representing the four different possible long-time behaviors:
chaotic (rule 00100, black circles), sponge-like (rule 00110, red
squares), segregated (rule 01011, green diamonds), and checkerboard
1x1 (rule 11010, blue triangles). Open symbols (connected by dashed
lines) correspond to system size $L=64$, while filled symbols correspond
to $L=128$. Averages taken over at least 1000 initial conditions.
Error bars are smaller than the symbols. Notice that for both rules
00100 and 00110 the curves for $L=64$ and $L=128$ are almost indistinguishable,
while finite-size effects are relevant for rules 01011 and 11010. }
\end{figure}
With respect to pattern formation, starting from random initial conditions,
the long-time behavior of the system strongly depends on the satisfaction
rule. The possible qualitatively distinct outcomes correspond to (i)
chaotic steady states; (ii) segregated states; (iii) checkerboard
1x1 states; (iv) sponge-like states. In what follows we discuss each
of these possible outcomes. With the exceptions of rules $01011$,
$10110$ and $11010$, which are unstable only around one of their
fixed points, those rules which are unstable around their fixed points
evolve to chaotic steady states, as described below. Examples of the
time dependence of the energy function and of the fraction of unsatisfied
agents for the distinct outcomes are shown in Fig. \ref{fig:ex_energ_dens}. 

\subsubsection*{Chaotic steady states}

\begin{table}
\begin{centering}
\begin{tabular}{|c|c|c|c|c|c|c|c|}
\hline 
Rule $s$ & 00000 & 00001 & 00010 & 00100 & 00101 & 01010 & 01001\tabularnewline
\hline 
$E^{\text{(sim)}}/L^{2}$ & $0.00$ & $-0.209$ & $-0.374$ & $0.00$ & $-0.219$ & $0.00$ & $+0.231$\tabularnewline
\hline 
$\rho_{u}^{\text{(sim)}}$ & $1.00$ & $0.895$ & 0.627 & $0.625$ & $0.524$ & $0.500$ & $0.587$\tabularnewline
\hline 
$\rho_{u}^{\text{(MF)}}$ & $1$ & $0.9375$ & $0.75$ & $0.625$ & $0.5625$ & $0.5$ & $0.6875$\tabularnewline
\hline 
\end{tabular}\medskip{}
\par\end{centering}
\begin{centering}
\begin{tabular}{|c|c|c|c|c|c|c|c|}
\hline 
Rule $s$ & 10001 & 10000 & 01000 & 11011 & 10100 & 10101 & 10010\tabularnewline
\hline 
$E^{\text{(sim)}}/L^{2}$ & $0.00$ & $+0.209$ & $+0.373$ & $\sim t/L^{2}$ & $+0.219$ & $0.00$ & $-0.230$\tabularnewline
\hline 
$\rho_{u}^{\text{(sim)}}$ & $0.875$ & $0.895$ & $0.627$ & $0.375$ & $0.524$ & $0.500$ & $0.586$\tabularnewline
\hline 
$\rho_{u}^{\text{(MF)}}$ & $0.875$ & $0.9375$ & $0.75$ & $0.375$ & $0.5625$ & $0.5$ & $0.6875$\tabularnewline
\hline 
\end{tabular}
\par\end{centering}
\caption{\label{tab:chaotic}The 14 rules leading to chaotic steady states
starting from random initial conditions. Simulation results for the
energy per agent $E^{\text{(sim)}}/L^{2}$ and the fraction $\rho_{u}^{\text{(sim)}}$
of unsatisfied agents correspond to long-time averages over 1000 initial
conditions. The uncertainty in the simulation results is in the last
digit. Results for the mean-field fraction $\rho_{u}^{\text{(MF)}}$
of unsatisfied agents have no uncertainty, and come from Eq. (\ref{eq:rhomf}).
See main text for a discussion on the behavior of the energy for rule
11011.}
\end{table}
There are 14 rules under which the long-time survival probability
is 100\% and no stable macroscopic domains of satisfied agents are
produced after a finite time in the thermodynamic limit. These rules
are listed in Table \ref{tab:chaotic}, along with the corresponding
stationary average total fraction of unsatisfied agents, $\rho_{u}$.
Since there are no macroscopic domains of satisfied agents, these
fractions can be compared with mean-field-like estimates obtained
by neglecting short-range correlations between agents. The reasoning
is based on counting the number of ways a given agent can have $k$
neighboring agents of its same type, and taking into account the corresponding
probabilities. As the total number of agents of both types is the
same, the steady-state fraction of unsatisfied agents (of both types)
is given by
\begin{equation}
\rho_{u}^{\text{(MF)}}\left(s\right)=\left(\frac{1}{2}\right)^{4}\sum_{k=0}^{4}\left(1-s_{k}\right)\frac{4!}{k!\left(4-k\right)!},\label{eq:rhomf}
\end{equation}
in which $s=s_{0}s_{1}s_{2}s_{3}s_{4}$ defines the rule. See the
Appendix for a derivation of the above result. 

On the other hand, this mean-field approximation, which is based on
the assumption that all configurations of the neighborhood of any
agent are equally probable, predicts that the average energy function
would be zero for all rules, and this is not generally compatible
with the square-lattice simulations. Table \ref{tab:chaotic} also
shows simulation results for the energy per agent $E/L^{2}$, as well
as a comparison between the mean-field estimates $\rho_{u}^{\text{(MF)}}$
and the simulation results $\rho_{u}^{\text{(sim)}}$ for the fraction
of unsatisfied agents. Notice that in general there is good agreement
between $\rho_{u}^{\text{(MF)}}$ and $\rho_{u}^{\text{(sim)}}$,
with a relative discrepancy below 15\%. For both $\rho_{u}^{\text{(MF)}}$
and $\rho_{u}^{\text{(sim)}}$, there is a symmetry in the steady-state
fraction of unsatisfied agents between a rule $s=s_{0}s_{1}s_{2}s_{3}s_{4}$
and its mirror rule $s^{\prime}=s_{4}s_{3}s_{2}s_{1}s_{0}$; see the
Appendix for a justification of this statement for any regular lattice.
Notice as well that, within numerical errors, $E\left(s^{\prime}\right)=-E\left(s\right)$,
if $s^{\prime}\neq s$, a result which is also justified in the Appendix.

\begin{figure}
\begin{centering}
\subfloat[]{\begin{centering}
\includegraphics[width=0.47\columnwidth]{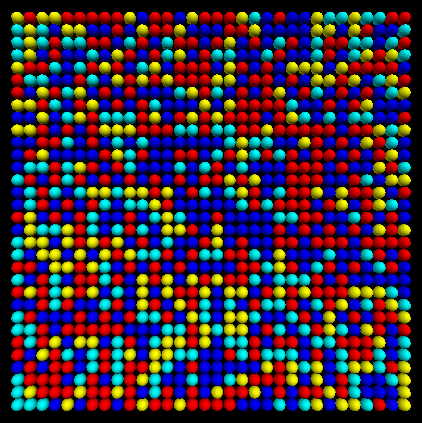}
\par\end{centering}
}\hfill{}\subfloat[]{\begin{centering}
\includegraphics[width=0.47\columnwidth]{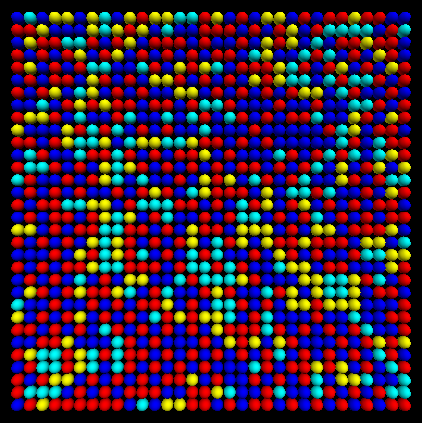}
\par\end{centering}
}
\par\end{centering}
\centering{}\subfloat[]{\begin{centering}
\includegraphics[width=0.47\columnwidth]{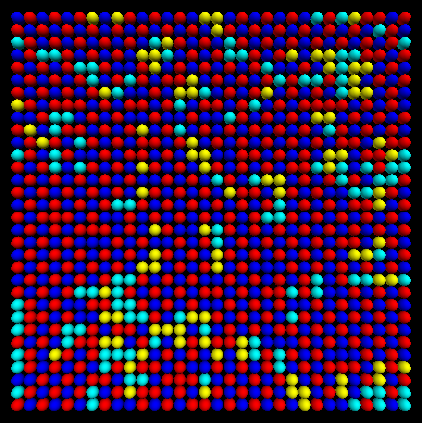}
\par\end{centering}
}\hfill{}\subfloat[]{\begin{centering}
\includegraphics[width=0.47\columnwidth]{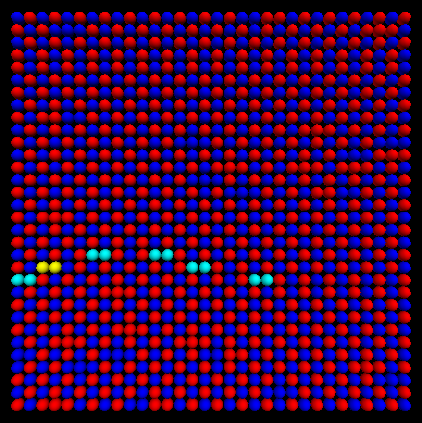}
\par\end{centering}
}\caption{\label{fig:11011_timeevol}Snapshots of the time evolution of the
system under rule 11011 and $L=32$. Unsatisfied agents are shown
in lighter colors (yellow for red agents, cyan for blue agents). (a)
The initial state, with energy per agent close to zero. (b) Configuration
when $E/L^{2}\simeq0.5$. (c) Configuration when $E/L^{2}\simeq1.1$.
(d) Configuration when $E/L^{2}=1.625$, the steady-state value for
that particular simulation, in which the unsatisfied agents are blinkers
(unsatisfied agents whose type alternates in time). Starting from
random initial conditions, the time required to reach this kind of
steady state diverges as $L\rightarrow\infty$.}
\end{figure}
\begin{figure}
\centering{}\subfloat[]{\begin{centering}
\includegraphics[width=0.98\columnwidth]{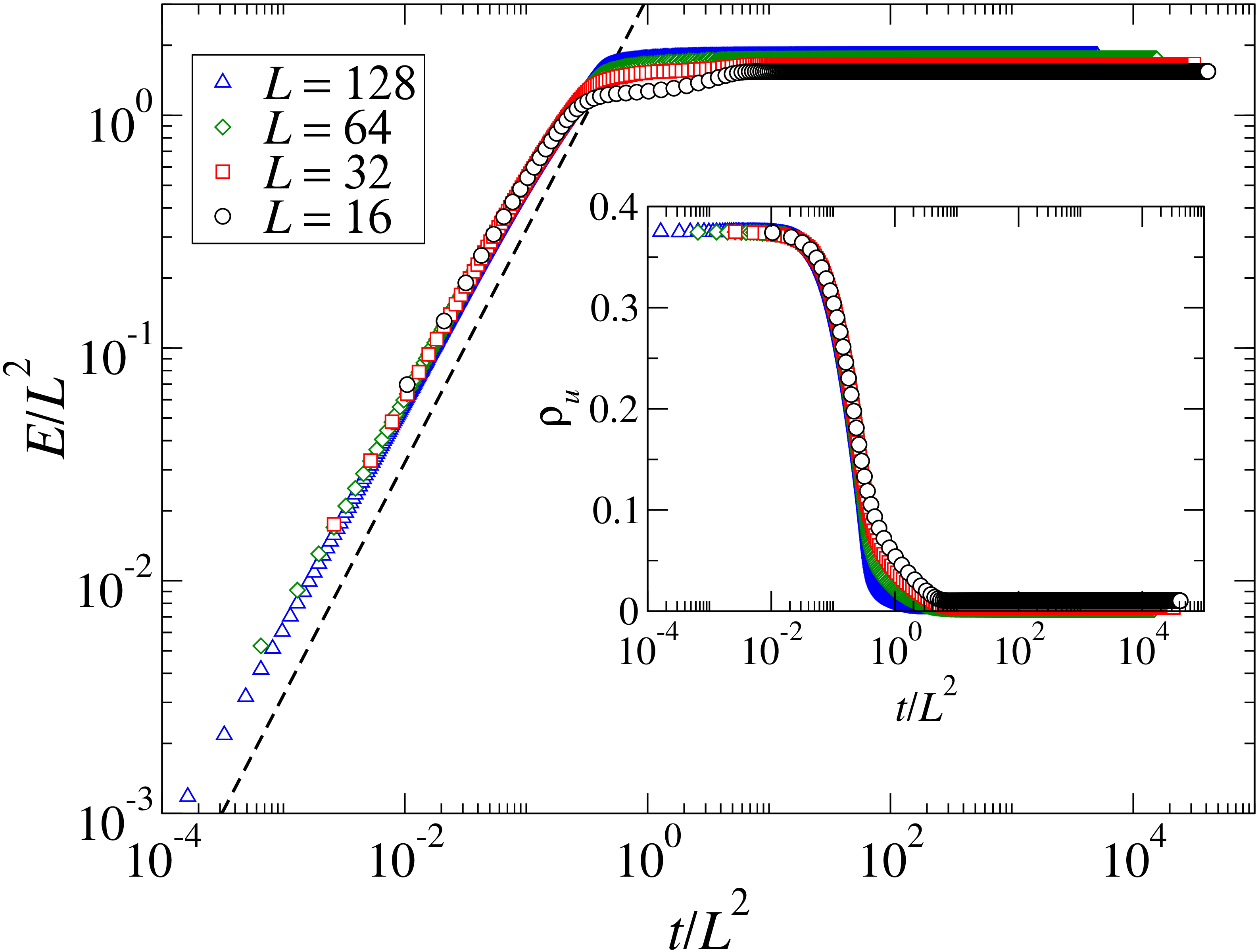}
\par\end{centering}
}\caption{\label{fig:Evst_11011}Main panel: energy per agent as a function
of time (rescaled by $L^{2}$) under rule 11011, for various system
sizes. The dashed line is proportional to $t/L^{2}$. Inset: fraction
of unsatisfied agents as a function of $t/L^{2}$. Error bars are
smaller than the symbols.}
\end{figure}
For mirror-symmetric rules, which are those for which $s^{\prime}=s$,
it is also clear from Table \ref{tab:chaotic} that the mean-field
predictions for $\rho_{u}$ and $E$ seem to be exact, which is related
to the fact that mirror-symmetry favors equal steady-state probabilities
$p_{k}$ and $p_{4-k}$ for neighborhood configurations of an unsatisfied
agent containing $k$ or $4-k$ agents of its opposite type, as well
as equal corresponding probabilities $q_{k}$ and $q_{4-k}$ for a
satisfied agent. For these pairs of configurations, the satisfaction
state of the agent is the same and the local contributions to the
energy function are $E_{k}$ and $E_{4-k}=-E_{k}$. However, in the
case of the energy function, rule $11011$ is an exception, due to
the fact that the stability of FP (1) under rule $11011$ is related
to an spontaneous symmetry breaking between $q_{k}$ and $q_{4-k}$
for $k\neq2$. Rule $11011$ is peculiar in that, for finite systems
of linear size $L$, a seemingly chaotic state, for which the average
energy nevertheless grows linearly with time, changes, after a timescale
which grows as $L^{2}$, to a checkerboard 1x1 pattern with defects.
Thus, in the thermodynamic limit the chaotic state is the only one
to be observed. Figure \ref{fig:11011_timeevol} illustrates the behavior
for a system with $L=32$, while Fig. \ref{fig:Evst_11011} shows
the time dependence of the average energy per agent and the average
fraction $\rho_{u}$ of unsatisfied agents for different system sizes.
As both the segregated state and the checkerboard 1x1 state are fixed
points of the rule, a random initial condition can be considered as
a mixture of checkerboard 1x1 and uniform domains of either type of
agent, separated by clusters of unsatisfied agents. Under the dynamics,
the average energy grows linearly with time, while $\rho_{u}$ remains
essentially constant, so that domains of satisfied agents forming
a local checkerboard 1x1 pattern grow in size by merging with each
other, quickly outcompeting uniform domains. This is associated with
the fact that, under this rule, agents are only unsatisfied if they
have exactly two neighbors of their same type. Direct inspection of
the possible local configurations makes it clear that the energy cannot
decrease under the dynamics, and in fact it only increases (always
by the same amount $\Delta E=6$) when two neighboring unsatisfied
agents (of opposite types) are interchanged. This initially happens
with a probability proportional to the inverse square of the number
of unsatisfied agents times the number of clusters of unsatisfied
agents, and therefore proportional to $L^{2}/L^{4}=L^{-2}$, so that
the average number of steps required for the energy to increase by
$\Delta E$ is proportional to $L^{2}$, which means a time interval
of order one. Thus, in order to reach the maximum allowed value of
the energy, which is proportional to $L^{2}$, a time of order $L^{2}$
is required. This can only be reached in finite systems.

\subsubsection*{Segregated states}

\begin{figure}
\begin{centering}
\subfloat[]{\begin{centering}
\includegraphics[width=0.47\columnwidth]{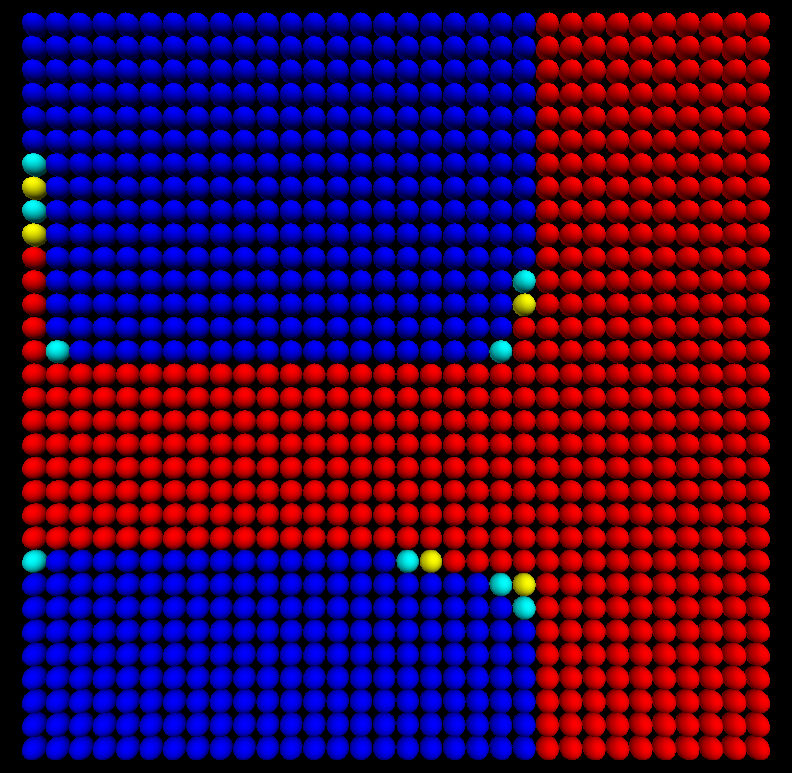}
\par\end{centering}
}\hfill{}\subfloat[]{\begin{centering}
\includegraphics[width=0.47\columnwidth]{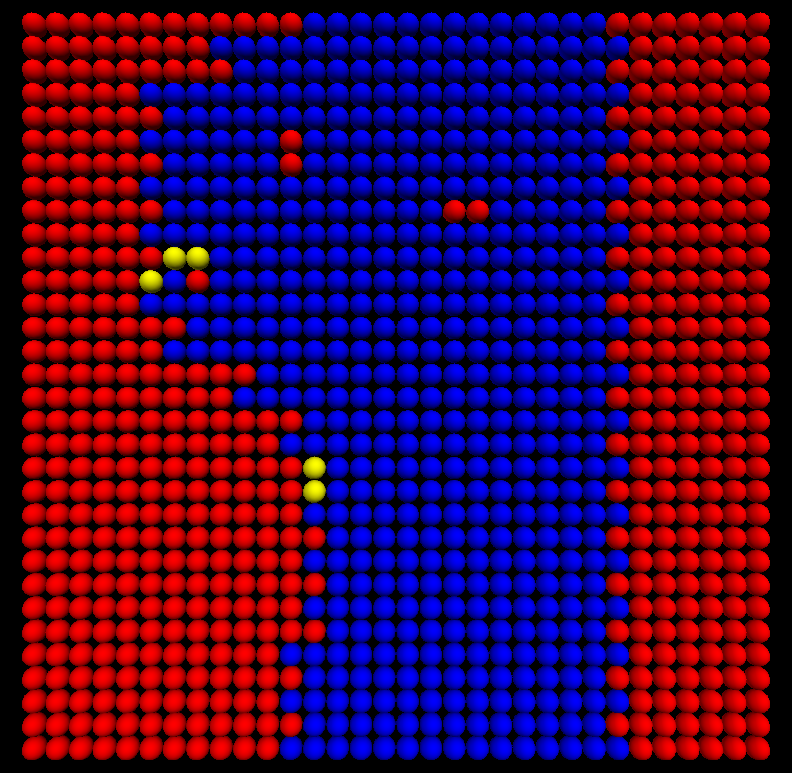}
\par\end{centering}
}
\par\end{centering}
\centering{}\subfloat[]{\begin{centering}
\includegraphics[width=0.47\columnwidth]{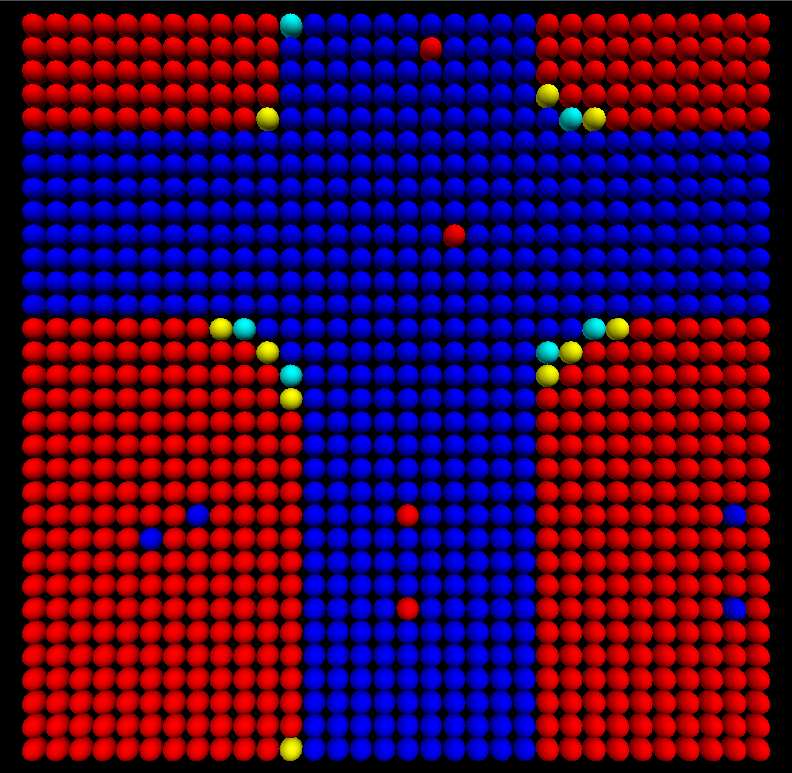}
\par\end{centering}
}\hfill{}\subfloat[]{\begin{centering}
\includegraphics[width=0.47\columnwidth]{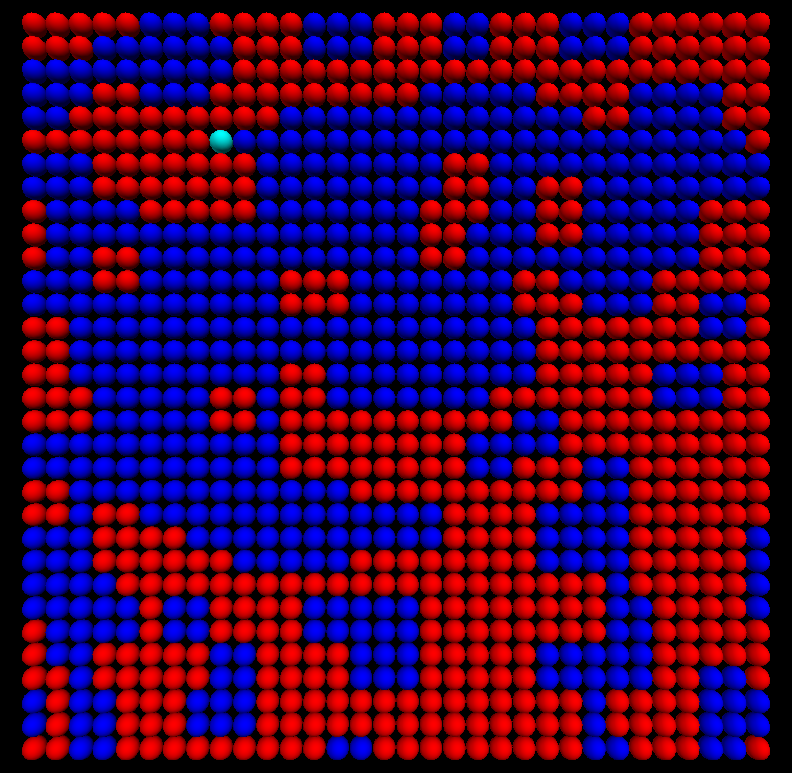}
\par\end{centering}
}\caption{\label{fig:segregated_patterns}Examples of long-time configurations
obtained for systems with $L=32$ under rules (a) 00011, (b) 01011,
(c) 10011, which give rise to nearly fully segregated configurations,
and (d) under rule 00111, for which the final pattern resembles a
wide sponge. Color code is the same as in Fig. \ref{fig:11011_timeevol}.}
\end{figure}
Under rules 00011, 01011 and 10011, the dynamics leads the system
to a segregated state with two domains, one for each type of agent,
possibly with a small number of ``defect'' static agents of the
opposite type. The domains are usually separated by a small fraction
of unsatisfied agents which goes to zero in the thermodynamic limit,
as they sit at the domain boundaries and therefore their number scales
at most with $L$, whereas the number of agents scales as $L^{2}$.
We therefore expect that the long-time average fraction of unsatisfied
agents $\rho_{u}$ decays as $1/L$, as long as the survival probability
$P_{s}$$\left(t\rightarrow\infty\right)$ is $100\%$. However, under
rule 00011 there is the possibility of forming perfectly linear domain
boundaries containing only satisfied agents, in which case the dynamic
freezes, so that $P_{s}\left(t\rightarrow\infty\right)$ is less than
unity. This becomes increasingly unlikely in the thermodynamic limit,
due to competition between horizontal and vertical boundaries, and
$P_{s}\left(t\rightarrow\infty\right)$ seems to approach unity as
$L\rightarrow\infty$, but rather slowly, leading to a slightly faster
decay of $\rho_{u}$ as $L^{-1.1}$. Figures \ref{fig:segregated_patterns}(a)-(c)
show examples of long-time configurations for the three segregating
rules.

For finite times and in the large $L$ regime, the fraction of unsatisfied
agents decays roughly as $\rho_{u}\left(t\right)\sim1/\sqrt{t}$ under
the three rules. This is somewhat surprising, as, except for rule
00011, in which agents have a strict preference for a neighborhood
containing a majority of agents of the same type, the other two rules
yielding segregated patterns are more tolerant to agents of the opposite
type, making an agent satisfied also if it has no neighbor (rule 10011)
or a single neighbor (rule 01011) of its same type. However, this
increased tolerance still leads to nearly fully segregated patterns
in the long run.

\subsubsection*{Checkerboard 1x1 states}

\begin{figure}
\begin{centering}
\subfloat[]{\begin{centering}
\includegraphics[width=0.47\columnwidth]{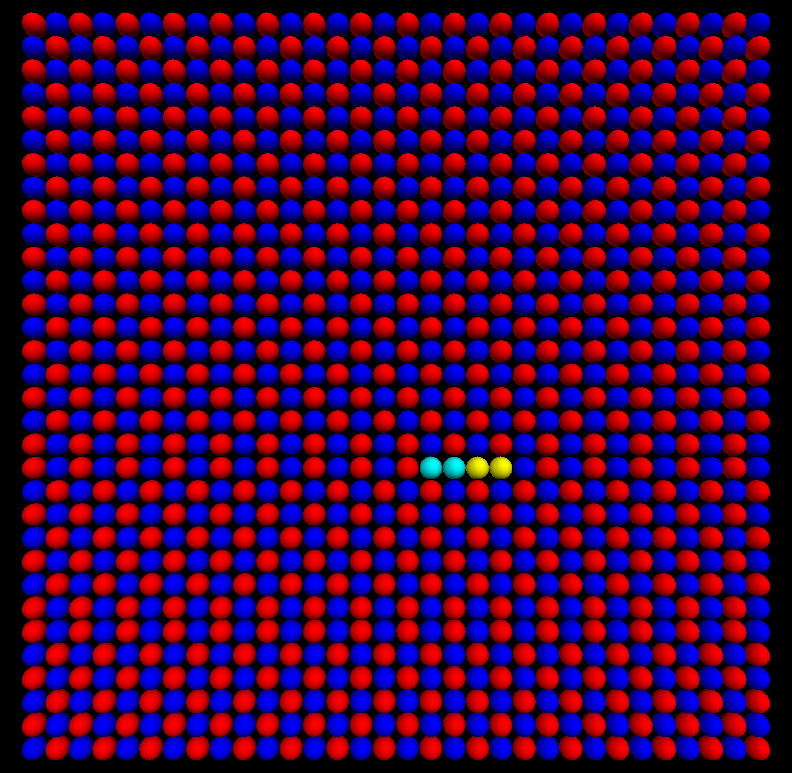}
\par\end{centering}
}\hfill{}\subfloat[]{\begin{centering}
\includegraphics[width=0.47\columnwidth]{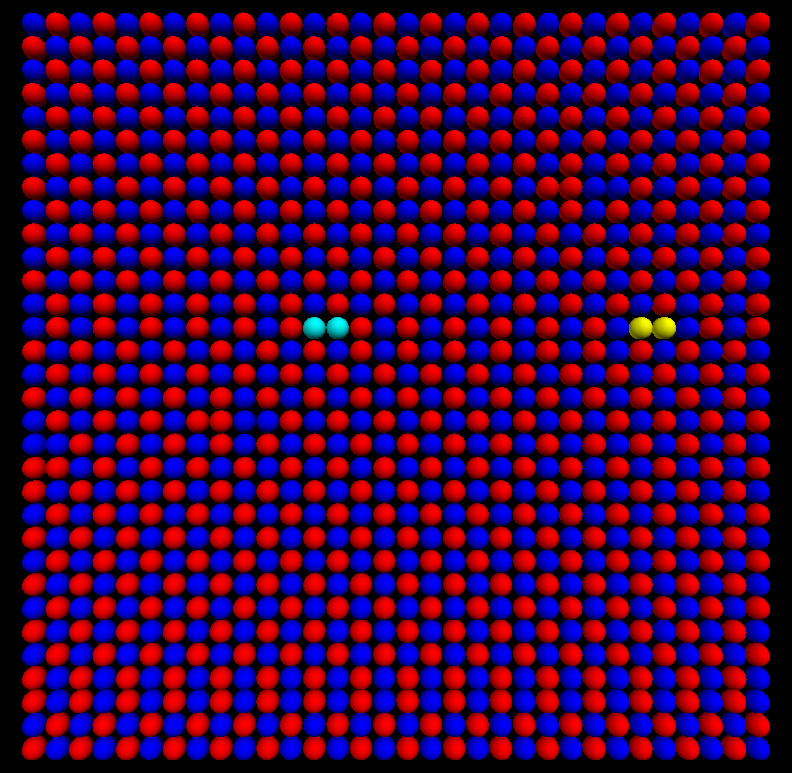}
\par\end{centering}
}
\par\end{centering}
\centering{}\subfloat[]{\begin{centering}
\includegraphics[width=0.47\columnwidth]{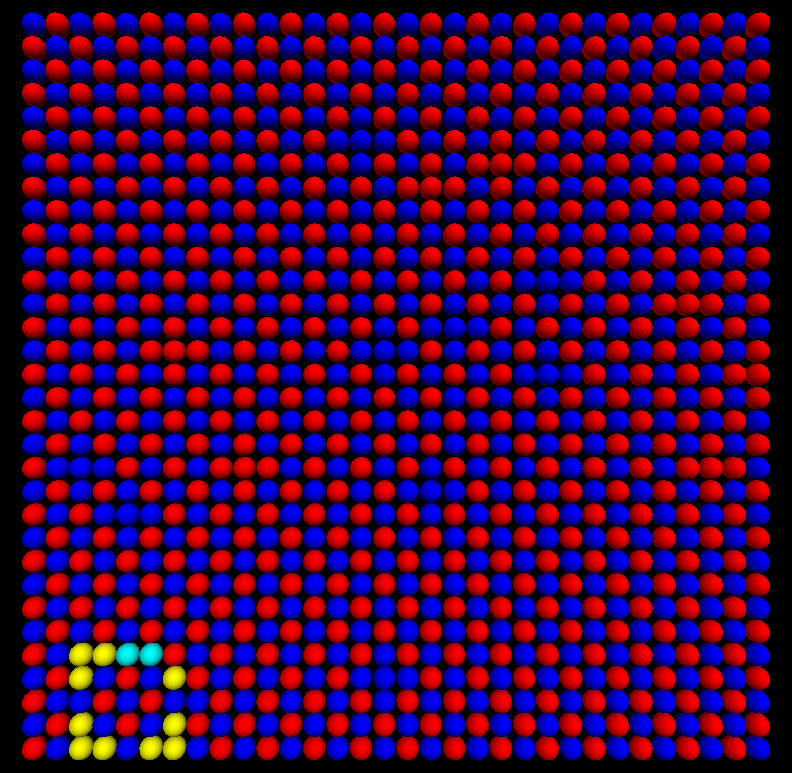}
\par\end{centering}
}\hfill{}\subfloat[]{\begin{centering}
\includegraphics[width=0.47\columnwidth]{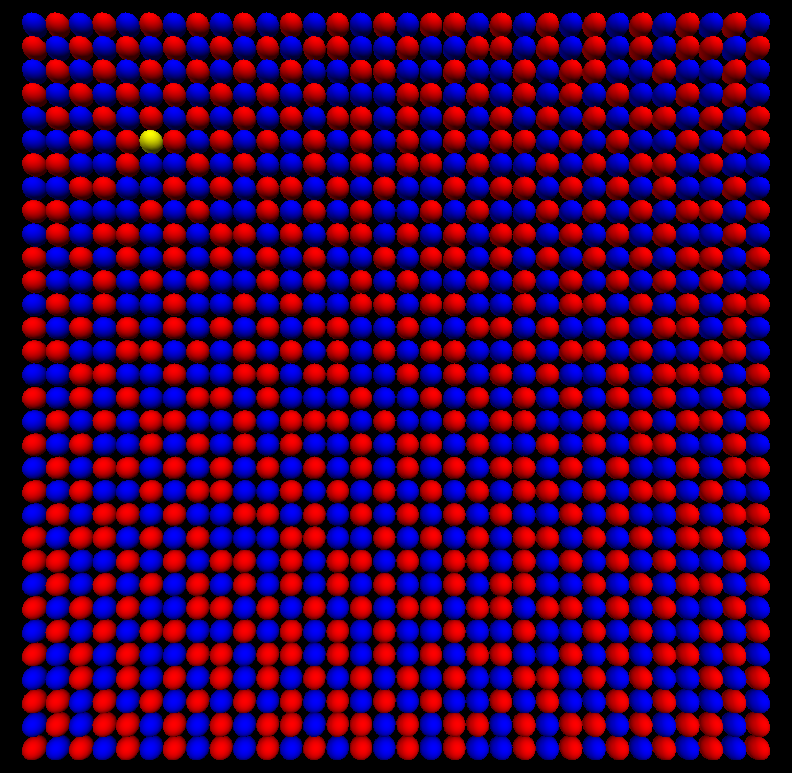}
\par\end{centering}
}\caption{\label{fig:checkerboard1x1_patterns}Examples of long-time configurations
obtained for systems with $L=32$ under rules (a) 11000, (b) 11010,
(c)11001, which give rise to anti-segregated patterns, and (d) under
rule 11100, for which the final pattern resembles a tight sponge.
Color code is the same as in Fig. \ref{fig:11011_timeevol}.}
\end{figure}
Under rules 11000, 11010 and 11001, which correspond to the mirror
rules of those yielding segregated states, the dynamics leads the
system to a checkerboard 1x1 state with a small number of defects
and a small average fraction of unsatisfied agents, which goes to
zero in the thermodynamic limit. Figure \ref{fig:checkerboard1x1_patterns}
illustrates the long-time patterns, which can also be described as
``anti-segregated''. 

The energy function is of course positive, as illustrated in Fig.
\ref{fig:ex_energ_dens}(a) for rule 11010. For the three rules we
numerically obtain the long-time behavior $\rho_{u}\sim1/L$, with
a small corresponding survival probability; see Fig. \ref{fig:ex_energ_dens}(b)
for rule 11010. The cusps seen in the curves are associated with the
characteristic time needed for the survival probability to start decreasing
from 100\%, which grows with $L^{2}$. The small number of long-time
configurations yielding blinkers is responsible for the asymptotic
values of $\rho_{u}$ and $E$, which are calculated as the average
of the corresponding values over the surviving simulations. Therefore,
the cusps are a statistical feature rather than a behavior observable
for a particular simulation.

For finite times and in the large $L$ regime, the fraction of unsatisfied
agents decays roughly as $\rho_{u}\left(t\right)\sim1/\sqrt{t}$ under
the three rules, as for their mirror rules.

\subsubsection*{Sponge-like states}

\begin{table}
\begin{centering}
\begin{tabular}{|c|c|c|c|}
\hline 
Rule & $E$ & LTB & $\rho_{u}$\tabularnewline
\hline 
\hline 
00111 & nonincreasing & frozen & $\sim L^{-1}$\tabularnewline
\hline 
11100 & nondecreasing & frozen & $\sim L^{-1}$\tabularnewline
\hline 
01111 & nonincreasing & frozen & $\sim L^{-1}$\tabularnewline
\hline 
11110 & nondecreasing & frozen & $\sim L^{-1}$\tabularnewline
\hline 
10111 & nonincreasing & frozen & $\sim L^{-1}$\tabularnewline
\hline 
11101 & nondecreasing & frozen & $\sim L^{-1}$\tabularnewline
\hline 
00110 & oscillating & blinkers & $4.1\%$\tabularnewline
\hline 
01100 & oscillating & blinkers & $4.1\%$\tabularnewline
\hline 
01110 & oscillating & blinkers & $12\%$\tabularnewline
\hline 
01101 & oscillating & mixed & $\sim L^{-1.2}$\tabularnewline
\hline 
10110 & oscillating & mixed & $\sim L^{-1.5}$\tabularnewline
\hline 
\end{tabular}
\par\end{centering}
\caption{\label{tab:labyrinth}The 11 rules under which the long-time arrangement
resembles a sponge, listing features of the energy function $E$,
the long-time behavior (LTB) and the asymptotic dependence of the
density $\rho_{u}$ of unsatisfied agents on the linear size $L$.}
\end{table}
\begin{figure}
\begin{centering}
\subfloat[]{\begin{centering}
\includegraphics[width=0.47\columnwidth]{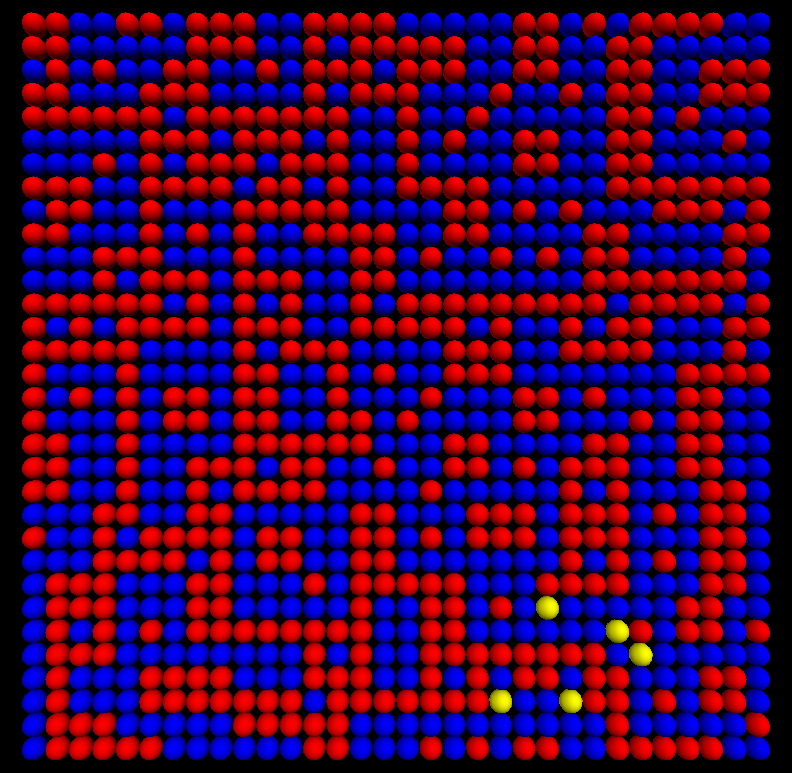}
\par\end{centering}
}\hfill{}\subfloat[]{\begin{centering}
\includegraphics[width=0.47\columnwidth]{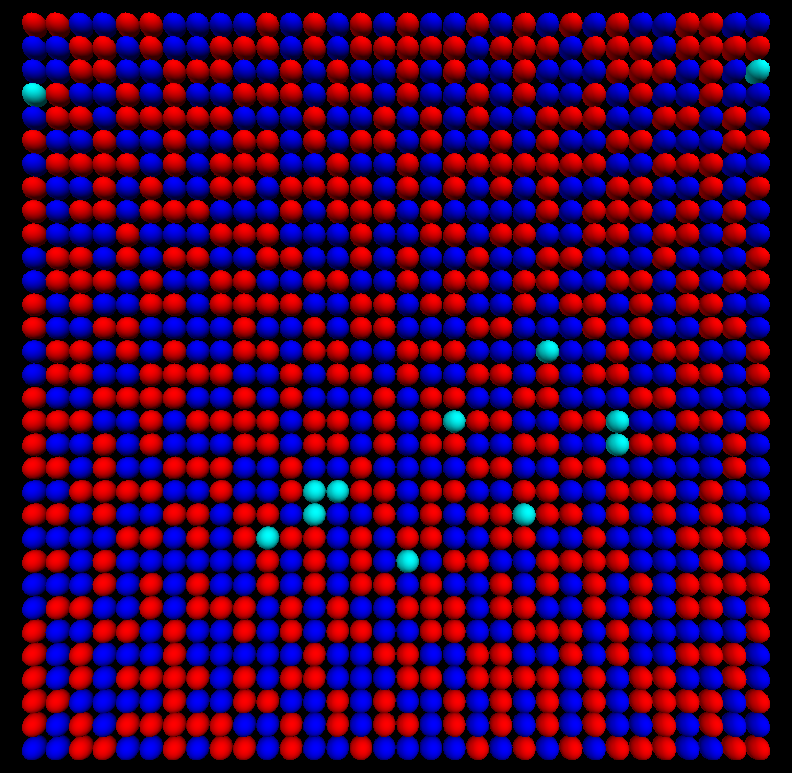}
\par\end{centering}
}
\par\end{centering}
\centering{}\subfloat[]{\begin{centering}
\includegraphics[width=0.47\columnwidth]{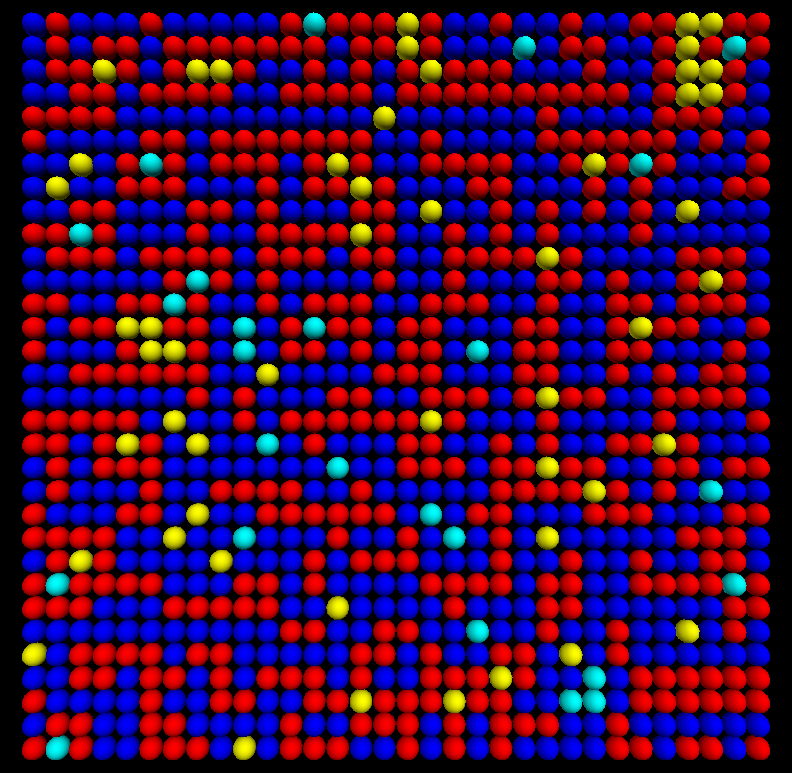}
\par\end{centering}
}\hfill{}\subfloat[]{\begin{centering}
\includegraphics[width=0.47\columnwidth]{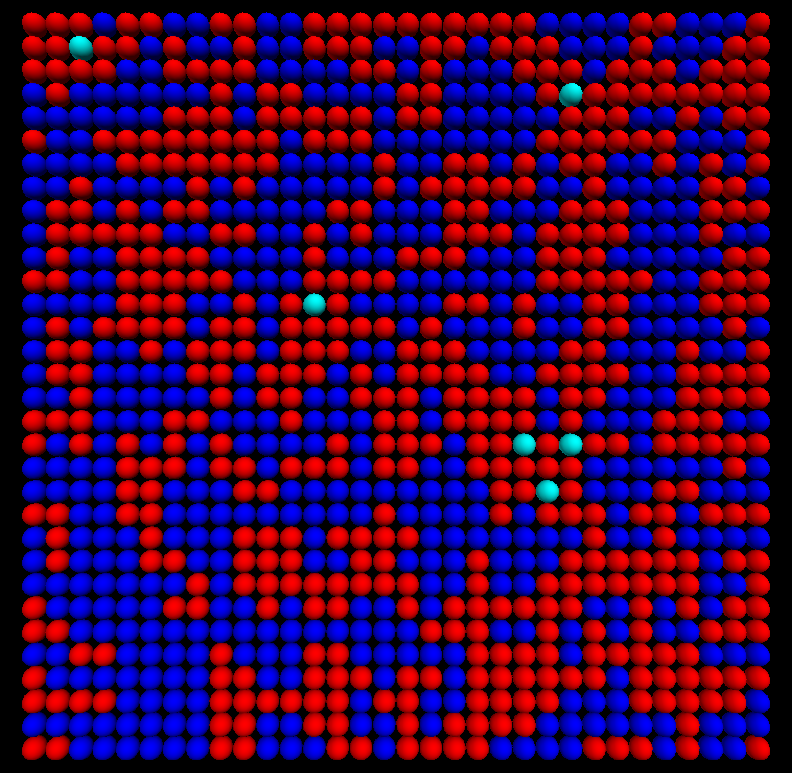}
\par\end{centering}
}\caption{\label{fig:labyrinth_patterns}Examples of sponge-like patterns obtained
for systems with $L=32$ under rules (a) 10110, (b) 01101, (c) 01110,
(d) 01111. Color code is the same as in Fig. \ref{fig:11011_timeevol}.}
\end{figure}
There are 11 rules under which the long-time behavior resembles a
sponge pattern, and which are listed in Table \ref{tab:labyrinth}.
All of these rules have the striped pattern and either the checkerboard
1x2 or the double-striped patterns as fixed points, and the sponge-like
long-time aspect can be attributed to the aggregation of domains of
either vertical or horizontal dimers or stripes. Examples of the patterns
are shown in Fig. \ref{fig:labyrinth_patterns}.

Under six of these rules the dynamics always leads to a frozen arrangement,
in which there are no unsatisfied agents of one type. This happens
for rules (see Table \ref{tab:labyrinth}) under which the energy
cannot decrease or cannot increase, and the average fraction of unsatisfied
agents upon freezing decreases as the inverse of the linear size $L$
of the system.

Under other rules (see Table \ref{tab:labyrinth}), the long-time
behavior is characterized by a small number of blinkers, whose density
remains finite in the thermodynamic limit, with a survival probability
of $100\%$. These are rules 01100 and 00110, related by a mirror
transformation, as well as the result of their combination, rule 01110.
Under rules 01100 and 00110, the fraction of unsatisfied agents decays
to its asymptotic value after a characteristic time of order 200,
exhibiting no $L$ dependence in the thermodynamic limit (see Fig.
\ref{fig:ex_energ_dens}). On the other hand, under rule 01110 this
decay happens after a time scale of order $L^{2}$, indicating that
it cannot be reached in the thermodynamic limit. The average fraction
$\rho_{u}=0.12$ indicated in Table \ref{tab:labyrinth} corresponds
to the finite-time value, and is quite close to the mean-field value
$\rho_{u}^{(\text{MF})}=0.125$ predicted by Eq. (\ref{eq:rhomf}).
The finite-time regime has no resemblance with a chaotic state, as
one is able to show by calculating the average fraction of initially
satisfied agents that ever become unsatisfied. This fraction also
turns out to be around 0.12, so that most local configurations, such
as those shown in Fig. \ref{fig:labyrinth_patterns}(c), are static.

Finally, there are two rules (01101 and its mirror rule 10110) under
which the long-time behavior strongly depends on the system size $L$.
The dynamics may either lead to a frozen state or to a small fraction
of blinkers, in both cases with a characteristic time corresponding
to only one MC step. For $L$ below about $128$, the long-time survival
probability (associated with the appearance of blinkers) is small,
reaches a minimum value around $L=128$ and increases rapidly above
$L$ around $256$, seeming to approach $100\%$ as $L\rightarrow\infty$.
The long-time average fraction of unsatisfied agents seems to decrease
with $L$ as a power law. We do not have an interpretation for this
odd finite-size behavior at this time.

Rule 00111 corresponds to the von-Neumann neighborhood version of
Schelling's initial model, in which an agent is satisfied if at least
half of its neighbors are of its same type. It yields a special type
of sponge-like pattern, with rather broad ``walls''. In this sense,
it is quite close to the segregated pattern. For comparison, we show
an example of the resulting patterns in Fig. \ref{fig:segregated_patterns}(d).

Rule 11100, corresponding to the mirror-transformed version of rule
00111, produces patterns reminiscent of the anti-segregated patterns,
despite the presence of small linear clusters of agents of the same
type. For comparison, we show an example of the resulting patterns
in Fig. \ref{fig:checkerboard1x1_patterns}(d).

\section{Conclusions}

In this paper we discussed an extension of Schelling's model on a
checkerboard, with no vacancies and the same number of agents of both
types, in which for each number $n$ of same-type nearest neighbors
we independently assign a binary satisfaction variable $s_{k}$ which
is equal to one only if the agent is satisfied with that condition,
and is equal to zero otherwise. Among the 32 resulting rules, 14 lead
to a chaotic steady state, one does not evolve dynamically, 11 give
rise to sponge-like patterns, while the remaining six rules lead to
nearly perfect segregation, in which almost all agents are surrounded
by agents of their same type, or to nearly perfect ``anti-segregation'',
in which almost all agents are surrounded by agents of the opposite
type.

The three rules leading to nearly perfect segregation share the fact
that agents are satisfied having either 3 or 4 neighbors of their
same type. However, both the rule under which agents are also satisfied
having only neighbors of the opposite type and the rule under which
agents are satisfied having a single neighbor of their same type also
lead to nearly perfect segregation. This is one more illustration
of the robustness of segregation induced by a mild preference for
a same-type neighborhood, already identified by Schelling.

The models studied here can be viewed as asynchronous cellular automata
subject to constant ``magnetization'', as we keep the total number
of agents of each kind fixed. This is in contrast with models subject
to the restriction of constant energy, as the Q2R automaton \cite{Vichniac1984,Pomeau1984},
which provides an efficient way to simulate the square-lattice Ising
model \cite{Herrmann1986,Herrmann1987}. The dynamical rule for this
automaton prescribes that spins are flipped, one sublattice at a time,
when they have two up and two down neighbors. This is similar to rule
11011, under which an agent is unsatisfied (and thus can move) only
when it has exactly two neighbors of its same kind. Besides keeping
the magnetization constant, the rule also differs from a fully asynchronous
version of the Q2R automaton in that neighboring agents, therefore
in different sublattices, can be switched in a single move, which
incidentally leads to the increase of the energy function.

A natural question to ask is how the observations described here depend
on the choice of neighborhood, on the assumption of an equal concentration
of agents of each type, and on the choice of equal satisfaction parameters
for both types of agents. We will provide an example of the effects
of relaxing those restrictions in a future publication. As an illustration
of the kind of behavior that appears, there occur transitions between
active and inactive phases as the relative concentration of agents
is varied, as in similar statistical-physics models \cite{Marro2005,Henkel2008},
such as the contact process and the voter model. This is of course
related to the existence of absorbing states represented by the fixed
points of the dynamics. Finally, another obvious extension of the
present work would be to study synchronous versions of the various
rules. 

\ack{}{This work was supported by the Brazilian agencies FUNCAP, CAPES,
CNPq, INCT-SC, INCT-FCx and FAPESP. EG acknowledges financial support
from Fondecyt 1190265, ECOS C16E01, STICSUD2018.}

\section*{Appendix}

Here we provide arguments pointing that, for any regular lattice,
rules related by mirror transformation give rise to the same stationary
density of unsatisfied agents, when there is the same number of agents
of both types. We also discuss the relation between the average energy
functions of a rule and of its mirror-transformed one, and derive
Eq. (\ref{eq:rhomf}).

Consider a regular lattice with coordination number $z$. Going beyond
the mean-field approximation, we can write the stationary fractions
of unsatisfied agents, $\rho_{u}\left(s\right)$, and of satisfied
agents, $\rho_{s}\left(s\right)$, as

\[
\rho_{u}\left(s\right)=\sum_{k=0}^{z}p_{k}\qquad\text{and}\qquad\rho_{s}\left(s\right)=\sum_{k=0}^{z}q_{k}
\]
in which $p_{k}$ ($q_{k}$) is the steady-state probability that,
under a given rule $s$, an unsatisfied (satisfied) agent has exactly
$k$ agents of its opposite type. 

Given a satisfaction rule defined by the parameter $s=s_{0}s_{1}s_{2}\cdots s_{z}$,
the corresponding mirror rule is defined by $s^{\prime}=s_{0}^{\prime}s_{1}^{\prime}s_{2}^{\prime}\cdots s_{z}^{\prime}$,
with $s_{k}^{\prime}=s_{z-k}$, and we represent by $p_{k}^{\prime}$
($q_{k}^{\prime}$) the steady-state probability that, under rule
$s^{\prime}$, an unsatisfied (satisfied) agent has exactly $k$ agents
of its opposite type. For each neighborhood configuration of an unsatisfied
agent of type $c$ containing $k$ agents of its opposite type under
rule $s$, applying the mirror transformation to the neighboring agents
generates a configuration containing $z-k$ agents of its opposite
type which makes the agent unsatisfied under rule $s^{\prime}$. Since
we assume an equal total number of agents of both types, and the dynamics
only involves unsatisfied agents, we expect $p_{z-k}^{\prime}=p_{k}$.
Therefore, 
\[
\rho_{u}\left(s^{\prime}\right)=\sum_{k^{\prime}=0}^{z}p_{k^{\prime}}^{\prime}=\sum_{k=0}^{z}p_{z-k}^{\prime}=\sum_{k=0}^{z}p_{k}=\rho_{u}\left(s\right).
\]
so that rules related by mirror transformation should give rise to
the same stationary density of unsatisfied agents, when there is the
same number of agents of both types.

Although the last result also implies the equality of the fraction
of satisfied agents under rules related by mirror transformation,
$\rho_{s}\left(s^{\prime}\right)=\rho_{s}\left(s\right)$, which yields
\[
\sum_{k=0}^{z}q_{k}=\sum_{k=0}^{z}q_{k}^{\prime},
\]
we cannot automatically conclude that $q_{z-k}^{\prime}=q_{k}$, as
that would imply the equivalence between each stable fixed point associated
with $s$ and a stable fixed point associated with $s^{\prime}$.
But in the square lattice, for instance, there is no equivalence between
the checkerboard 1x1 pattern, in which each agent is surrounded by
four agents of the opposite type, and the fully segregated pattern,
which consists of two uniform domains separated by a boundary along
which an agent necessarily has at least one neighbor of the opposite
type. This is reflected in the fact that rule $11011$, which is mirror-symmetric,
has a very small $q_{0}$ but a $q_{4}$ close to $1$, as shown by
simulation results.

The average stationary energy per agent under rule $s$ is given by
\[
\overline{E}\left(s\right)=\frac{1}{2}\sum_{k=0}^{z}E_{k}\left(p_{k}+q_{k}\right).
\]
Under the mirror rule, we obtain
\[
\overline{E}\left(s^{\prime}\right)=\frac{1}{2}\sum_{k^{\prime}=0}^{z}E_{k^{\prime}}\left(p_{k^{\prime}}^{\prime}+q_{k^{\prime}}^{\prime}\right)=\frac{1}{2}\sum_{k=0}^{z}E_{z-k}\left(p_{z-k}^{\prime}+q_{z-k}^{\prime}\right).
\]
But $p_{z-k}^{\prime}=p_{k}$, and from the definition of the energy
we have $E_{k}=-E_{z-k}$, so that 
\[
\overline{E}\left(s^{\prime}\right)=-\frac{1}{2}\sum_{k=0}^{z}E_{k}\left(p_{k}+q_{z-k}^{\prime}\right).
\]
If we can assume that $q_{z-k}^{\prime}=q_{k}$, as in the cases in
which there are no stable regular fixed points, then we can conclude
that $\overline{E}\left(s^{\prime}\right)=-\overline{E}\left(s\right)$.
This is what is numerically verified for 13 of the 14 rules in Table
\ref{tab:chaotic}, but not for rule $11011$, under which checkerboard
1x1 domains outcompete the uniform domains in the long-time dynamics.

The mean-field approximation assumes that all local neighborhood configurations
of an unsatisfied agent containing $k$ agents of its opposite type
are equally probable, which is a reasonable assumption as long as
there are no stable regular fixed points of the dynamics. Taking the
number of agents of both types as equal, this leads to 
\[
p_{k}=\left(1-s_{k}\right)\frac{z!}{k!\left(z-k\right)!}\left(\frac{1}{2}\right)^{k}\left(\frac{1}{2}\right)^{z-k}=\left(\frac{1}{2}\right)^{z}\left(1-s_{k}\right)\frac{z!}{k!\left(z-k\right)!},
\]
as in Eq. (\ref{eq:rhomf}).\bigskip{}

\bibstyle{iopart-num}\bibliographystyle{/Users/apvieira/Dropbox/latex/sty/iop/iopart-num}
\bibliography{agents}

\end{document}